# Method for real-time measurement of the nonlinear refractive index

Manuel P. Fernández,[1,2,3] Laureano A. Bulus Rossini, [1,2,3] Pablo A. Costanzo Caso[1,2,3]

[1]*LIAT - Laboratorio de Investigación Aplicada en Telecomunicaciones (CNEA, Comisión Nacional de Energía Atómica), Av. Bustillo 9500, Bariloche 8400 (RN), Argentina*

[2]*CONICET, CCT Patagonia Norte, Bariloche 8400 (RN), Argentina*

[3]*Instituto Balseiro (UNCuyo-CNEA), Bariloche 8400 (RN), Argentina*

[*]*Corresponding author: manuel.fernandez@ib.edu.ar; pcostanzo@ib.edu.ar*

In this work, we propose a novel method for continuous real-time measurement of the dynamics of the nonlinear refractive index $n_2$. This is particularly important for characterizing phenomena or materials (such as biological tissues, gases and other compounds) whose nonlinear behavior or structure varies rapidly with time. The proposed method ingeniously employs two powerful tools: the spectral broadening induced by self-phase modulation and the real-time spectral analysis using the dispersive Fourier transformation. The feasibility of the technique is experimentally demonstrated, achieving high-speed measurements at rates of several MHz.

Third-order nonlinear interaction between light and materials (known as the Kerr effect) is a major research field [1]. Measurements of the nonlinear refractive index $n_2$ are of utmost interest for many applications, including the study of the nonlinear properties of novel materials [2], biological tissues [3]-[4], liquids [5] and gases [6]. Moreover, nonlinear waveguides such as highly nonlinear optical fibers and integrated devices are gaining much research interest as photonic sensors [7]-[8].

The most extended method for measuring $n_2$ of thin samples is the Z-scan method [9], which is based on the self-focusing effect of a probe beam and the power measurement behind an aperture versus the sample position. A temporal analog of the Z-scan substitutes the translation stage with a tunable dispersive line [10], allowing to characterize both bulk materials and waveguides. In order to obtain the nonlinear coefficient of waveguides such as optical fibers, several techniques based on self-phase modulation (SPM), cross-phase modulation (XPM) and four-wave mixing (FWM) have been developed, which employ interferometric methods [11]-[12] that require time-consuming power scans and are sensitive to polarization effects. An approach for measuring the nonlinearity of very short fibers based on the acousto-optic interaction effect was recently demonstrated [13]. Techniques based on measuring the SPM-broadened spectrum are widely used to characterize the nonlinearity of bulk samples [2] and waveguides [14], [15], but they rely on slow optical spectrum analyzers or tunable filters, and thus, they are not suitable for high-speed measurements. All of the above-mentioned methods require performing parametric scans, e.g. sample position or input power, which is a major drawback for some applications, as the process become time and energy consuming. Single-shot measurements are also highly desired in order to avoid damage on the sample under test. To our knowledge, the techniques reported in [5] and [16] are the only non-scanning methods that allow the $n_2$ measurement in bulk samples, but they require complex pump-probe setups.

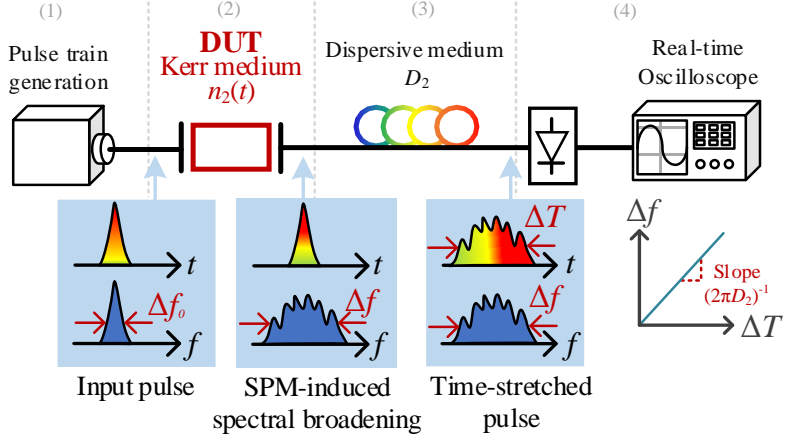

Fig. 1. Principle of the system for real-time measurement of a time-varying nonlinear refractive index $n_2(t)$. The lower waveforms represent the evolution of a single probe-pulse. DUT: device under.

In this Letter, we propose for the first time to our knowledge, a novel method for real-time tracking of the nonlinear refractive index in a nonlinear Kerr medium at rates as high as hundreds of MHz. This technique exploits the linear mapping of the SPM-broadened probe-pulse spectrum into its temporal waveform when it propagates through a highly dispersive medium. The non-interferometric and non-scanning nature of the system makes it an extremely simple and robust tool for ultra-fast measurements of a time-varying nonlinearity.

The principle of the proposed $n_2$ measurement system is illustrated in Fig. 1, and it comprises four subsystems: (1) a pulsed optical source, (2) the device under test (DUT), which is a Kerr nonlinear material, (3) a highly dispersive medium and (4) a detection system. The optical source emits a train of probe short pulses at a given repetition rate, which propagate through the nonlinear DUT. For each pulse, the SPM-induced peak nonlinear phase shift is given by

$$\phi_{NL} = \frac{2\pi}{\lambda} \frac{n_2}{A_{eff}} P_0 L_{eff}, \qquad (1)$$

where $P_0$ is the input pulse peak power, $\lambda$ is the central wavelength, $A_{eff}$ and $L_{eff}$ are the effective area and effective propagation length of the DUT, respectively, and $n_2$ is the nonlinear refractive index. The time-dependent phase induced by SPM generates new frequencies and broadens the pulse spectrum. Therefore, the spectral width at the output of the DUT, $\Delta f$, is a function of the peak nonlinear phase shift, such that $\Delta f = f(\phi_{NL})$. This function can be either theoretically or experimentally determined.

After probing the DUT, the pulses propagate in the linear regime through a second-order dispersive material, which has a high chromatic dispersion parameter $D_2$. If the dispersion is sufficiently high, such that the far-field temporal Fraunhofer condition is accomplished, the pulses spectrum are mapped into their temporal waveform [17]-[18]. In such a case, the equation that relates the temporal width of the dispersed pulses $\Delta T$ to their spectral width can be written as

$$\Delta T = 2\pi D_2 \Delta f \quad . \tag{2}$$

Finally, the resulting time-stretched pulses are detected using a single photodiode and digitized by a real-time oscilloscope.

The procedure for real-time measurement of the nonlinear coefficient can be summarized as follows. After the probe pulses propagate through the DUT, the highly dispersive medium maps their spectrum into their temporal waveforms. Then, the output signal is detected and sampled, and by means of a pulse-by-pulse digital signal processing, the temporal width of the time-stretched pulses is measured. The spectral width corresponding to each pulse is then derived according to Eq. (2) and the nonlinear phase shift is recovered applying the inverse function $\phi_{NL} = f^{-1}(\Delta f)$. Therefore, by a proper characterization of the DUT ($A_{eff}$ and $L_{eff}$) and the experimental setup ($P_0$ and $\lambda$), the nonlinear refractive index can be obtained from each pulse by means of Eq. (1). Therefore, this method allows for single-shot and ultra-fast measurements at speeds as high as the laser repetition rate (up to hundreds of MHz).

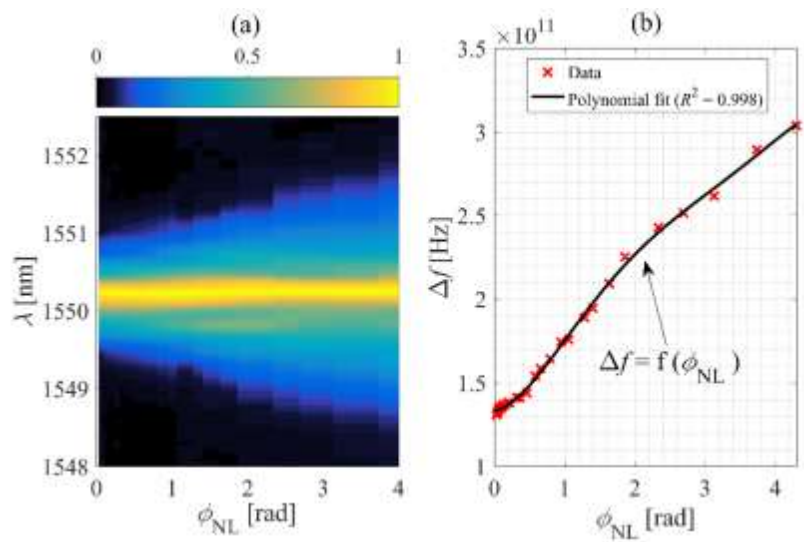

Fig. 2. (a) Measured evolution of the spectrum with $\phi_{NL}$. (b) Spectral width versus the nonlinear phase shift. The calibration function is obtained by means of a fifth-order polynomial fitting of the data.

In order to demonstrate the measurement principle, we implemented an experimental setup based on the schematic of Fig. 1. As optical source, we used a pulsed laser (Onefive Katana 15), which delivers high peak-power pulses of width $T\sim30$ ps centered at 1550.2 nm at a tunable repetition rate. The dispersive medium consists of a dispersion compensation module (DCM) for telecommunications networks with dispersion parameter $D_2 = 2123$ ps$^2$. The dispersed pulses are detected by a photodetector (Newport 1544-A) and digitized using a real-time oscilloscope (R&S-RTO-1044).

Prior to the experiments, we conducted a calibration procedure to experimentally obtain the relation $\Delta f = f(\phi_{NL})$. For this purpose, instead of a DUT we placed a calibration fiber with known parameters, and measured the output spectrum as a function of $\phi_{NL}$, as illustrated in Fig. 2(a). From these measurements, we obtained the function $\Delta f = f(\phi_{NL})$ by means of a polynomial fitting of the measured data, as shown in Fig. 2(b). The spectral width $\Delta f$ is measured at the 20% of the amplitude maximum, which accounts for 95% of the input pulse energy, and similarly, the width $\Delta T$ of the time-stretched pulses is measured at the 20% of their maximum.

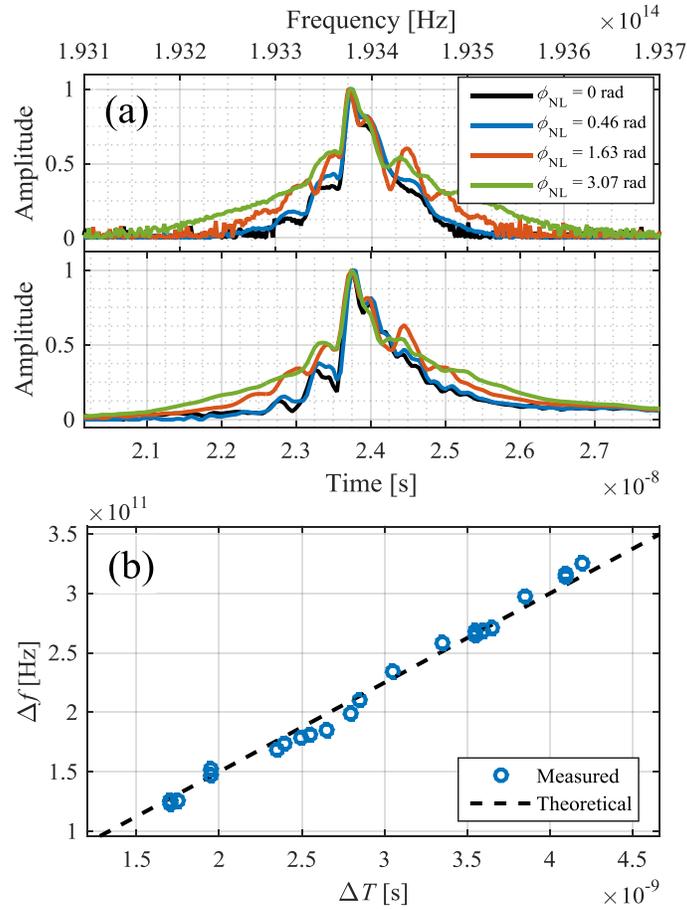

Fig. 3. Dispersive Fourier transformation of the spectrally broadened pulses. (a) Pulse spectrum at the output of the DUT. (b) Pulse temporal waveform at the output of the dispersive medium. (c) Measured spectral width versus temporal width.

The frequency-to-time mapping of the spectrally broadened pulses after propagation in the dispersive medium is shown in Fig. 3 for different values of the nonlinear phase shift. Figure 3(a) shows the spectrum at the output of the DUT measured using an optical spectrum analyzer (upper waveforms) and the pulses detected at the output of the DCM (lower waveforms). It is clearly appreciated how the pulse spectrum is mapped into its temporal waveform, and thus broader spectrums yield wider detected pulses. The linear frequency-to-time mapping relation can be clearly appreciated in Fig. 3(b), where it is represented the measured spectral width versus the temporal width at the DCM output for several $\phi_{NL}$. It can be seen that the experimental data agrees very well with the theoretically expected relation described by Eq. (2).

In order to demonstrate the feasibility of the method to perform fast dynamic measurements of $n_2$, we conducted a proof-of-concept experiment to measure a dynamic nonlinear phase shift $\phi_{NL}(t)$ in a DUT. In this case, the DUT is a single-mode fiber with $n_2 = 2.6 \times 10^{-20}$ m$^2$/W, $A_{eff} = 80$ μm$^2$, and $L_{eff} = 986$ m. By considering that both $n_2(t)$ and $P_0(t)$ produce the same effect on $\phi_{NL}(t)$ in terms of SPM-induced spectral broadening (see Eq. (1)), we emulate the nonlinear refractive index variation through a time-varying input peak power.

The generated nonlinear phase shift is the hopping sequence shown in Fig. 4(a), which has level transitions following the sequence [0.53, 1.35, 1.63, 2.33] rad at intervals of 666.66 ns. Simultaneously, in the right axis of Fig. 4(a), it is shown the equivalent $n_2(t)$ evolution considering that the DUT is probed by a constant input power $P_0(t) = 1$ W, which emulates the $n_2$ sequence [1.07, 2.7, 3.2, 4.6]x10$^{-20}$ m$^2$/W. The pulse repetition rate of the laser source is set to 30 MHz (a repetition period of 33.33 ns), and thus the induced $\phi_{NL}$ remains constant over 20 probe pulses and then step up to the new sequence value for the next 20 pulses. The normalized detected signal is depicted in Fig. 4(b), where the appreciated intensity fluctuations are mainly due to noise associated to the detection stage. In order to mitigate this source of uncertainty when measuring the pulse width, it is performed the time-averaging of $N_{avg}$ consecutive pulses prior to measuring $\Delta T$, which significantly increase the signal-to-noise ratio (SNR). For instance, uncorrelated Gaussian noise increases the SNR by a factor $N_{avg}$.

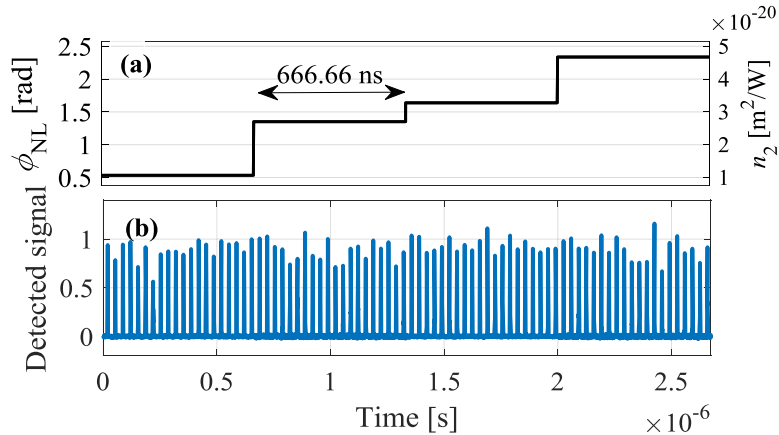

Fig. 4. (a) Generated time-varying nonlinear phase shift and equivalent $n_2$ assuming $P_0(t) = 1$ W, $A_{eff} = 80$ μm$^2$, $L_{eff} = 986$ m. (b) Detected signal normalized by the received power of each interval.

The result of measuring $\phi_{NL}$ (and hence $n_2$) from the detected signal is shown in Fig. 5, where different numbers of time-averaged successive pulses are considered. When no averaging is performed and the pulse width is directly measured from each single pulse, the estimated $\phi_{NL}(t)$ is a relatively noisy signal following the theoretically expected behavior, as seen in Fig. 5(a). When an averaging of $N_{avg}$ = 2, 5 and 10 successive pulses is performed, the measurement error is greatly reduced, and the measured $\phi_{NL}$ rapidly approaches its real value, as shown in Fig. 5(b)-(d). Since $\phi_{NL}$ is calculated from $N_{avg}$ pulses, the SNR enhancement occurs at the expense of a reduction of $N_{avg}$ times the temporal acquisition resolution: while for no averaging, the acquisition rate is equal to the pulse repetition rate (30 MHz), for $N_{avg}$ = 2, 5, 10, the effective acquisition rates are 15 MHz, 6 MHz and 3 MHz, respectively. Therefore, the number of averages should result from a trade-off between measurement accuracy and measurement speed.

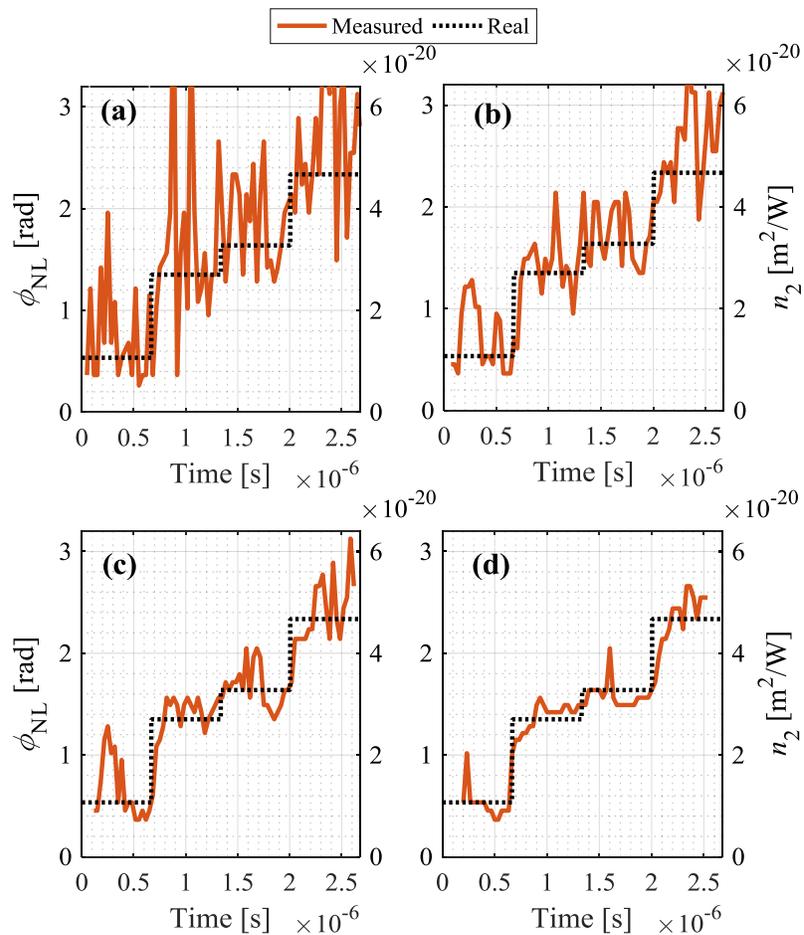

Fig. 5. Measured nonlinear phase shift and the equivalent $n_2$ for different number of averaged consecutive pulses. (a) No averaging; (b) $N_{avg}$ = 2; (c) $N_{avg}$ = 5; (d) $N_{avg}$ = 10.

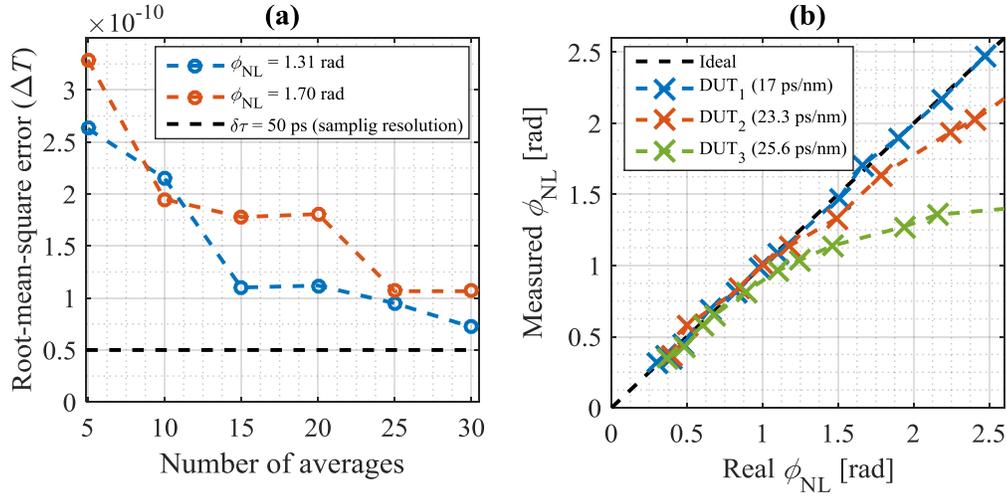

Fig. 6. (a) Measured RMSE of the pulse width versus the number of averaged successive pulses. (b) Effects of chromatic dispersion on the estimation of the nonlinear phase shift.

**Noise effects.** The SNR depends on the experimental setup conditions, i.e. optical power arriving at the photodiode, detector's noise and laser intensity noise. As we have qualitatively seen, time-averaging pulses lead to an improved measurement accuracy, but reduces the acquisition rate. To investigate this effect, we measured the root-mean-square error (RMSE) of the temporal width $\Delta T$ from 200 pulses, where each of these pulses results from averaging $N_{avg}$ pulses. Figure 6(a) depicts the RMSE as a function of $N_{avg}$ when $\phi_{NL}$=1.31 rad and $\phi_{NL}$=1.70 rad. In both cases, the mean input power at the detector was adjusted to be −37 dBm, yielding a pessimistic scenario in which the SNR is relatively poor. It can be seen that the RMSE rapidly decreases and approaches the limit of the oscilloscope sampling resolution (50 ps) even for moderate number of averages.

**Dispersion considerations.** In the general case, SPM dominates over dispersion effects in the DUT for propagation distances $L$ and input pulse power levels such that $(\gamma P_0)^{-1} < L \ll T^2/|\beta_2|$, were $\beta_2$ is the group velocity dispersion and $\gamma$ is the nonlinear coefficient defined as $2\pi n_2/(\lambda A_{eff})$. If this requirement is not accomplished, dispersion effects on the output spectrum should be properly considered. To this respect, we compared the measured $\phi_{NL}$ for three devices with different and non-negligible dispersion parameters. For instance, $DUT_1$, $DUT_2$ and $DUT_3$ have an accumulated dispersion of 17 ps/nm, 23.3 ps/nm and 25.6 ps/nm, respectively. Figure 6(b) depicts the measured versus the real generated $\phi_{NL}$, which is that defined by Eq. (1). As it can be seen, for $DUT_1$ there is a perfect match between the real and measured parameter, since dispersion effects on the probe pulses are negligible within the considered measurement range ($\phi_{NL} < 2.5$ rad). For $DUT_2$ and $DUT_3$, the measured and real $\phi_{NL}$ agree for smaller nonlinear phase shifts (in this case $\phi_{NL} < 1$ rad). For larger values there exists an underestimation on the measured parameter since dispersion leads to a non-negligible pulse broadening in the time domain, which results in a lower peak power and, consequently, a reduced spectral broadening. However, this drawback can be overcome by taking into account the interplay of both nonlinearity and dispersion on the function $\Delta f = f(\phi_{NL})$, for example, by means of numerical simulations, as it is commonly done in the standard SPM-based measurement technique [14]. Moreover, the input power of the probe pulses can be chosen such that the system operates in the region for which dispersion effects are negligible ($\phi_{NL} < 1$ rad in the example of Fig. 8).

**Resolution.** We finally discuss the nonlinear phase shift resolution, $\delta\phi_{NL}$, as it ultimately determines the minimum measurable $n_2$ variation. In this system, the spectral resolution $\delta f$ is related to the sampling resolution $\delta\tau$ and the dispersion parameter $D_2$ by means of Eq. (2). In our setup, since the sampling period is $\delta\tau = 50$ ps, the time-domain spectral resolution is found to be $\delta f = 50 \times 10^{-12}/(2\pi \times 2123 \times 10^{-24}) = 3.75$ GHz. The parameters $\delta f$ and $\delta\phi_{NL}$ are related through the slope of the function $\Delta f = f(\phi_{NL})$. For instance, in the interval $\phi_{NL} \in [0.58$ rad, $2.32$ rad] this function is well described by a linear equation with slope $5.22 \times 10^{10}$ Hz/rad. Therefore, within this interval, the nonlinear phase shift resolution is found to be $\delta\phi_{NL} = 0.072$ rad.

**Conclusions.** We have proposed a novel technique that allows for real-time and high-speed measurement of the SPM-induced nonlinear phase shift in nonlinear Kerr materials (both bulk samples and waveguides). Therefore, parameters related to $\phi_{NL}$, such as the nonlinear refractive index, can be measured at high rates without performing any type of parametric scan. In this method, the spectral broadening of the probe pulses, associated to $n_2$, is directly measured in the time-domain by means of the real-time Fourier transformation of the short probe pulses when they propagate in a highly dispersive medium. Accurate and ultra-fast measurements at rates of several MHz were experimentally demonstrated.

To our knowledge, this is the first time that such continuous ultra-fast measurements of the nonlinearity are demonstrated. Therefore, the presented technique has potential applications in many novel and high impact fields such as medicine, biology, materials science and chemical sensing. It also constitutes an interesting approach for the development of novel high-speed sensing schemes, in which the nonlinear properties of the sensing element, i.e. the DUT, are modified by the parameter of interest.

**Acknowledgments.** This paper was partially supported by CNEA, CONICET, UNCUYO and Sofrecom Argentina SA. MPF is fellow of CONICET and PACC and LABR are professors at IB and researchers of CONICET.-

**REFERENCES**

[1] B. Gu, C. Zhao, A. Baev, K. Yong, S. Wen, and P. N. Prasad, Adv. Opt. Photon. **8**, 2 (2016).

[2] H. Qian, X. Yuzhe, and L. Zhaowei, Nature Commun. **7**, 13153 (2016).

[3] C. T. Nabeshima, S. I. P. Alves, A. M. F. Neto, F. R. O. Silva, R. E. Samad, and L. C. Courrol, *Frontiers in Optics.* Paper JTh2A.132 *(2016).*

[4] A. Bezryadina, T. Hansson, R. Gautam, B. Wetzel, G. Siggins, A. Kalmbach, J. Lamstein, D. Gallardo, E. J. Carpenter, A. Ichimura, R. Morandotti, and Z. Chen, Phys. Rev. Lett. **119**, 058101 (2017).

[5] S. S. Farahani, K. Madanipour, and A. Koohian, Appl. Opt. **56**, 13 (2017).

[6] Á. Börzsönyi, Z. Heiner, A. Kovács, M. Kalashnikov, and K. Osvay, Opt. Expr. **18**, 25 (2010).

[7] J. C. Knight and D. V. Skryabin, Opt. Expr. **15**, 23 (2007)

[8] J. Leuthold, C. Koos, and W. Freude, Nature Phot. **4**, 535–544 (2010).

[9] M. Sheik-bahae, A. A. Said, and E. W. Van Stryland, Opt. Lett. **14**, 17 (1989).

[10] F. Louradour, E. Lopez-Lago, V. Couderc, V. Messager, and A. Barthelemy, Opt. Lett. **24**, 19 (1999).

[11] D. Monzón-Hernández, A. N. Starodumov, Y. O. Barmenkov, I. Torres-Gómez and F. Mendoza-Santoyo, Opt. Lett. **23**, 16 (1998).

[12] K. Li, Z. Xiong, G. Peng, and P. Chu, Opt. Commun. **136**, 3-4 (1997).

[13] E. Rivera–Pérez, A. Carrascosa, A. Díez, E. Alcusa-Sáez and M. Andrés, Appl. Phys. Lett., **113,** 1 (2018).

[14] R. Stolen, W. Reed, K. Kim, and G. Harvey, J. Ligthw. Technol. **16**, 6 (1998).

[15] K. S. Kim, R. H. Stolen, W. A. Reed, and K. W. Quoi, Opt. Lett. **19**, 4 (1994).

[16] I. Dancus, S. Popescu, and A. Petris, Opt. Expr. **21**, 25 (2013).

[17] K. Goda, and B. Jalali, Nature Phot. **7**, 102-112 (2013).

[18] J. Azaña and M. Muriel, J. Quant. Electr. **36**, 5 (2000).

# FULL REFERENCES

1. B. Gu, C. Zhao, A. Baev, K. Yong, S. Wen, and P. N. Prasad, "Molecular nonlinear optics: recent advances and applications," Adv. Opt. Photon. **8**, 328-369 (2016)

2. H. Qian, X. Yuzhe, and L. Zhaowei, "Giant Kerr response of ultrathin gold films from quantum size effect." *Nature Communications*, 7 (2016).

3. C. T. Nabeshima, S. I. P. Alves, A. M. F. Neto, F. R. O. Silva, R. E. Samad, and L. C. Courrol, "Z-scan technique: A new concept for Diagnosis of Prostate Cancer in blood," in *Frontiers in Optics 2016*, OSA Technical Digest (online) (Optical Society of America, 2016), paper JTh2A.132.

4. A. Bezryadina, T. Hansson, R. Gautam, B. Wetzel, G. Siggins, A. Kalmbach, J. Lamstein, D. Gallardo, E. J. Carpenter, A. Ichimura, R. Morandotti, and Z. Chen, "Nonlinear self-action of light through biological suspensions", Phys. Rev. Lett. **119**, 058101 (2017).

5. S. Farahani, K. Madanipour, and A. Koohian Nonscanning Moiré deflectometry for measurement of nonlinear refractive index and absorption coefficient of liquids. *Applied Optics*, 56(13), (2017).

6. Börzsönyi, Á., Heiner, Z., Kovács, A., Kalashnikov, M. and Osvay, K. "Measurement of pressure dependent nonlinear refractive index of inert gases" *Optics Express*, 18(25), (2010)

7. J. C. Knight and D. V. Skryabin, "Nonlinear waveguide optics and photonic crystal fibers," Opt. Express **15**, 15365-15376 (2007)

8. J. Leuthold, C. Koos, and W. Freude, "Nonlinear silicon photonics", Nature photonics, 4 (2010).

9. M. Sheik-bahae, A. A. Said, and E. W. van Stryland, "High sensitivity, single-beam n2 measurement," Opt. Lett. 14, 955–957 (1989).

10. Louradour, F., Lopez-Lago, E., Couderc, V., Messager, V. and Barthelemy, A. (1999). "Dispersive-scan measurement of the fast component of the third-order nonlinearity of bulk materials and waveguides." *Optics Letters*, 24(19), p.1361

11. D. Monzón-Hernández, A. N. Starodumov, Y. O. Barmenkov, I. Torres-Gómez, and F. Mendoza-Santoyo, "Continuous-wave measurement of the fiber nonlinear refractive index," *Optics Letters*, vol. 23, no. 16, p. 1274, 1998.

12. K. Li, Z. Xiong, G. Peng, and P. Chu, "Direct measurement of nonlinear refractive index with an all-fibre Sagnac interferometer," *Optics Communications*, vol. 136, no. 3-4, pp. 223–226, 1997.

13. Rivera–Pérez, E., Carrascosa, A., Díez, A., Alcusa-Sáez, E. and Andrés, M. (2018). An approach to the measurement of the nonlinear refractive index of very short lengths of optical fibers. *Applied Physics Letters*, 113(1), p.011108.

14. R. Stolen, W. Reed, K. Kim, and G. Harvey, "Measurement of the nonlinear refractive index of long dispersion-shifted fibers by self-phase modulation at 1.55 μm," *Journal of Lightwave Technology*, vol. 16, no. 6, pp. 1006–1012, 1998

15. K. S. Kim, R. H. Stolen, W. A. Reed, and K. W. Quoi, "Measurement of the nonlinear index of silica-core and dispersion-shifted fibers," Opt. Lett. **19**, 257-259 (1994).

16. I. Dancus, S. Popescu, and A. Petris, "Single shot interferometric method for measuring the nonlinear refractive index." *Optics Express*, 21(25), (2013)

17. K. Goda and B. Jalali, "Dispersive Fourier transformation for fast continuous single-shot measurements," *Nature Photonics*, vol. 7, no. 2, pp. 102–112, 2013.

18. J. Azaña and M. Muriel, "Real-time optical spectrum analysis based on the time-space duality in chirped fiber gratings," *IEEE Journal of Quantum Electronics*, vol. 36, no. 5, pp. 517–526, 2000.